\documentclass[twocolumn,showpacs,preprintnumbers,amsmath,amssymb]{revtex4}

\usepackage{graphicx}
\usepackage{dcolumn}
\usepackage{bm}

\begin{document}

\title{Quantifying Fluctuation/Correlation Effects on the Order-Disorder Transition of Symmetric Diblock Copolymers}

\author{Jing Zong and Qiang Wang}
\email{q.wang@colostate.edu}
\affiliation{Department of Chemical and Biological Engineering, Colorado State University, Fort Collins, CO 80523-1370}

\date{\today}

\begin{abstract}
Using fast off-lattice Monte Carlo simulations with experimentally accessible fluctuations, we report the first systematic study unambiguously quantifying the shift of the order-disorder transition (ODT) $\chi^*$ of symmetric diblock copolymers from the mean-field prediction $\chi^*_{\rm MF}$. Our simulations are performed in a canonical ensemble with variable box lengths to eliminate the restriction of periodic boundary conditions on the lamellar period. Exactly the same model system (Hamiltonian) is used in both our simulations and mean-field theory; the ODT shift is therefore due to the fluctuations/correlations neglected by the latter. While $\chi^* / \chi^*_{\rm MF} - 1 \propto \bar{\mathcal{N}}^{-k}$ is found with $\bar{\mathcal{N}}$ denoting the invariant degree of polymerization, $k$ decreases around the $\bar{\mathcal{N}}$-value corresponding to the close packing of polymer segments as hard spheres, indicating the short-range correlation effects.
\end{abstract}

\maketitle

\begin{figure}[b]
\begin{center}
\includegraphics[height=5.1cm]{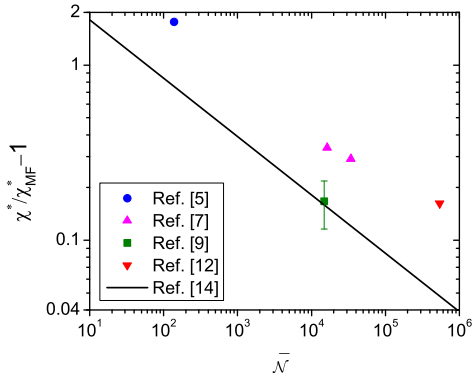}
\end{center}
\caption{\label{Liter}(Color online) Some literature results on the ODT shift. See text for details.}
\end{figure}
Owing to the well developed polymer self-consistent field (SCF) calculations\cite{GlenBK}, good or even quantitative understanding has been achieved for the self-assembly of flexible linear diblock copolymer (DBC) melts in bulk\cite{SCFDBC}. Due to its mean-field approximation, however, SCF theory gives qualitatively incorrect predictions in the region near the order-disorder transition (ODT) where the system fluctuations it neglects become important. In particular, it fails to capture the fluctuation-induced first-order phase transition for ODT of symmetric DBC\cite{FlucBF} and the direct transition between the gyroid and disordered phases\cite{Schulz}. Although the former is a classic problem in polymer science and has been extensively studied by experiments, theories and simulations, no \emph{quantitative} understanding of the fluctuation effects on ODT is achieved. Fig.~\ref{Liter} summarizes recent simulation results on the ODT shift of symmetric DBC from the mean-field prediction vs. the invariant degree of polymerization $\bar{\mathcal{N}} \equiv n R_{e,0}^3 / V$, where $n$ denotes the number of copolymer chains, $R_{e,0}$ the end-to-end distance of an ideal chain, and $V$ the system volume; these are the most accurate data obtained using each method as explained below.

Beardsley and Matsen performed conventional Monte Carlo (MC) simulations on a face-centered cubic (FCC) lattice in a canonical ensemble with replica-exchange to study symmetric DBC of $N=30$ segments modeled by the self- and mutual-avoiding walks with nearest-neighbor repulsion between A and B segments $\epsilon_{\rm AB}$ (in units of $k_B T$, where $k_B$ is the Boltzmann constant and $T$ the thermodynamic temperature); about 20\% lattice sites were unoccupied, treated as an athermal solvent.\cite{MatsMC3} From the peak of constant-volume heat capacity, they determined ODT to be $\chi^* N \equiv z \epsilon^*_{\rm AB} N = 40.5$, where $z=12$ is the lattice coordination number.\cite{MatsMC3} The mean-field ODT was determined to be $\chi_{\rm MF}^* N = 14.654$ using lattice SCF calculations based on the same Hamiltonian as used in their simulations.\cite{MatsMC1}

Using fast off-lattice Monte Carlo (FOMC) simulations in an isothermal-isobaric ensemble, de Pablo and co-workers studied compressible symmetric DBC melts modeled by discrete Gaussian chains (DGC) of $N=64$ with a position-independent but anisotropic pair potential of cubic symmetry.\cite{Detch3} They determined $\chi^* N$ by calculating the chemical potentials of the disordered and lamellar phases.\cite{Detch3} Note that their $\chi^* N$ was renormalized to take into account the short-range correlations (due to the finite interaction range used in their simulations) approximately; their $\chi^*_{\rm MF} N$ was therefore determined using the random-phase approximation for DGC with Dirac $\delta$-function interactions.\cite{Detch3,Note1}

M\"uller and Daoulas performed single-chain-in-mean-field (SCMF) simulations to study compressible symmetric DBC melts modeled by DGC of $N=32$, and determined $\chi^* N = 13.65 \pm 0.1$ by comparing the free energies of the disordered and lamellar phases.\cite{Daou2} Their SCMF simulation is similar to canonical-ensemble FOMC simulation with a spatial discretization scheme (i.e., with a position-dependent and anisotropic pair potential)\cite{FOMC}, except that a second-order term in the energy difference due to MC trial moves is neglected by the quasi-instantaneous field approximation\cite{Daou1}. They also estimated $\chi^*_{\rm MF} N = 11.7 \pm 0.5$ by fitting the lamellar composition profiles (averaged over directions parallel to the lamellar interfaces) of a system at $\bar{\mathcal{N}} \approx 2.5 \times 10^9$ at various $\chi N$ to a sinusoidal function and then extrapolating the square of the so-obtained amplitude of the composition profiles to 0.\cite{Daou2}

Using field-theoretic simulations in a canonical ensemble, Fredrickson and co-workers studied incompressible DBC melts modeled by continuous Gaussian chains with Dirac $\delta$-function interactions.\cite{FTSFene} They determined $\chi^* N = 12.2$ (for slightly asymmetric DBC with a composition $f=0.49$) by comparing the free energies of the disordered and lamellar phases.\cite{FTSFene} For this ``standard'' model (which cannot be directly used/tested in molecular simulations), $\chi^*_{\rm MF} N = 10.495$ for $f=0.5$ is the well-known Leibler's result\cite{Leib} and is used in Fig.~\ref{Liter}.

Finally, the prediction of the fluctuation theory\cite{FlucFH} by Fredrickson and Helfand (FH), $\chi^*/\chi^*_{\rm MF} - 1 = 3.91 \bar{\mathcal{N}}^{-1/3}$, is also shown as a straight line in Fig.~\ref{Liter}. This theory is based on the Ohta-Kawasaki effective Hamiltonian\cite{OK} for the above ``standard'' model and the Hartree analysis by Brazovskii\cite{Braz}. Although the latter is rigorously accurate only for $\bar{\mathcal{N}} \gtrsim 10^{10}$\cite{FlucFH}, the FH prediction has often been compared with simulation results at much smaller $\bar{\mathcal{N}}$ in the literature. With the different models and methods used and the scarce simulation data in Fig.~\ref{Liter}, it is clear that the ODT shift of symmetric DBC due to fluctuations is a problem far from being well understood.

\begin{figure}[t]
\begin{center}
\includegraphics[height=5.4cm]{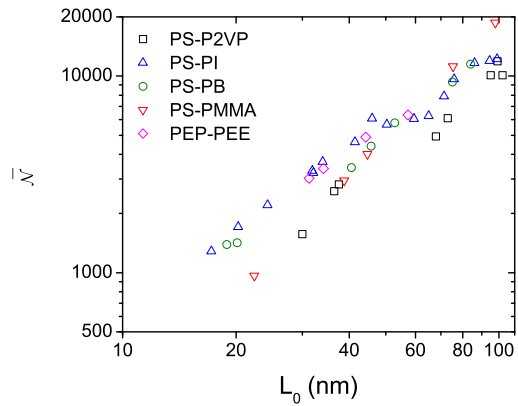}
\end{center}
\caption{\label{L0}(Color online) The invariant degree of polymerization $\bar{\mathcal{N}}$ vs. the bulk lamellar period $L_0$ for nearly symmetric DBC melts commonly used in experiments.\cite{L0Exp,Note2}}
\end{figure}

The system fluctuations are controlled by $\bar{\mathcal{N}}$, as suggested by the FH theory. For monodisperse DBC melts where each copolymer chain consists of $N_{\rm A}$ monomers of type A followed by $N_{\rm B}$ monomers of type B, with $R_{e,0}^2 = N_{\rm A} a_{\rm A}^2 + N_{\rm B} a_{\rm B}^2$ and $n N = \bar{\rho}_0 V$, we have $\bar{\mathcal{N}} = N \bar{\rho}_0^2 \bar{a}^6$; here $N = N_{\rm A} + N_{\rm B}$, $\bar{\rho}_0 \equiv \phi_{\rm A} \rho_{0, \rm A} + (1 - \phi_{\rm A}) \rho_{0, \rm B}$, $\bar{a} \equiv \sqrt{f a_{\rm A}^2 + (1-f) a_{\rm B}^2}$, $f \equiv N_{\rm A} / N$, $a_{\rm A}$ ($a_{\rm B}$) denotes the statistical segment length of the A (B) block, $\rho_{0, \rm A}$ ($\rho_{0, \rm B}$) the A (B) monomer number density, and $\phi_{\rm A}$ the overall volume fraction of the A block. Fig.~\ref{L0} shows $\bar{\mathcal{N}}$ vs. the bulk lamellar period $L_0$ for nearly symmetric DBC melts commonly used in experiments\cite{L0Exp}; the range of $L_0 = 10\sim 100$nm, which is of interest for most applications of DBC, roughly corresponds to $\bar{\mathcal{N}} = 500 \sim 20,000$\cite{Note2}.

In all conventional molecular simulations with hard excluded-volume interactions (e.g., the Lennard-Jones potential or the self- and mutual-avoiding walks), $\bar{\mathcal{N}}$ is on the same order of magnitude as $N$, which is at most one hundred or so for concentrated polymer solutions or melts (e.g., the leftmost data point in Fig.~\ref{Liter} is at $\bar{\mathcal{N}} \approx 139$). Using soft potentials that allow particle overlapping is therefore the \emph{only} way to study DBC melts within the above $\bar{\mathcal{N}}$-range (i.e., with experimentally accessible fluctuations) at present, where $n N = \bar{\rho}_0 V$ no longer holds and $N$ becomes a chain discretization parameter that does not correspond to the actual chain length used in experiments. This point is crucial for understanding coarse-grained models with soft potentials, the use of which is the basic idea of the recently proposed fast MC simulations\cite{FOMC,FLMC}.

In this work, we perform extensive FOMC simulations to quantify the ODT shift of symmetric DBC melts with the following Hamiltonian: $\mathcal{H} = \mathcal{H}^C + \mathcal{H}^E$, where $\mathcal{H}^C = (3 k_B T / 2 a^2) \sum_{k=1}^n \sum_{s=1}^{N-1} \left({\bf R}_{k,s+1} - {\bf R}_{k,s}\right)^2$ is the Hamiltonian due to chain connectivity, and $\mathcal{H}^E = (1/2 \kappa \rho_0) \int{\rm d}{\bf r}{\rm d}{\bf r}' \left[\hat{\rho}_{\rm A}({\bf r}) + \hat{\rho}_{\rm B}({\bf r})\right] u_0(|{\bf r}-{\bf r}'|) \left[\hat{\rho}_{\rm A}({\bf r}') + \hat{\rho}_{\rm B}({\bf r}')\right] + (\chi / \rho_0) \int{\rm d}{\bf r}{\rm d}{\bf r}' \hat{\rho}_{\rm A}({\bf r}) u_0(|{\bf r}-{\bf r}'|) \hat{\rho}_{\rm B}({\bf r}')$ is the Hamiltonian due to non-bonded interactions; here $a$ is the effective bond length, ${\bf R}_{k,s}$ denotes the spatial position of the $s^{\rm th}$ segment on the $k^{\rm th}$ chain, $\kappa$ and $\chi$ are the generalized Helfand compressibility\cite{Helf} for the copolymer melts and the generalized Flory-Huggins interaction parameter for the repulsion between A and B segments, respectively, $\rho_0 \equiv n N / V$, $\hat{\rho}_{\rm A}({\bf r}) \equiv \sum_{k=1}^n \sum_{s=1}^{N_{\rm A}} \delta({\bf r} - {\bf R}_{k,s})$ and $\hat{\rho}_{\rm B}({\bf r}) \equiv \sum_{k=1}^n \sum_{s=N_{\rm A}+1}^N \delta({\bf r} - {\bf R}_{k,s})$ are the microscopic number density of A and B segments at spatial position ${\bf r}$, respectively, and $u_0(r) = (15 k_B T / 2 \pi \sigma^3) (1-r/\sigma)^2$ for $r<\sigma$ and 0 otherwise is a normalized (i.e., $\int{\rm d}{\bf r} u_0(|{\bf r}|) = k_B T$) isotropic pair potential depending only on the distance $r$ between two segments with $\sigma$ denoting the finite interaction range. Note that $u_0(r)$ is essentially the same as the potential for the conservative force used in dissipative particle dynamics\cite{DPD}; its Fourier transform is positive definite, thus avoiding the formation of clustered crystals at large $\rho_0$\cite{CluCry}. Taking $R_{e,0} = \sqrt{N-1} a$ as the length scale, we have five parameters in our model: $\bar{\mathcal{N}}$, $\chi N$, $N/\kappa$, $N$, and $\sigma / a$.

Our simulations are performed in a canonical ensemble with trial moves of random hopping, reptation, pivot\cite{PVTOL}, and box-length change. Note that the highly efficient pivot algorithm cannot be used in multi-chain simulations with hard excluded-volume interactions due to its extremely small acceptance rates. With soft potentials, however, we can achieve an acceptance rate of about 70\% here. We also use replica-exchange\cite{RE} at different $\chi N$ (with acceptance rates of 50$\sim$80\%) to further improve our sampling efficiency.

In simulations of periodic structures such as lamellae, the periodic boundary conditions limit the allowed orientations of the structure and thus its period. In particular, for lamellae with a normal direction ${\bf n}$ in a simulation box with length $L_j$ in the ${\bf j} (={\bf x},{\bf y},{\bf z})$ direction, $L_j {\bf j} \bm{\cdot} {\bf n} = n_j L({\bf n})$ must be satisfied, where $n_j$ is the number of periods contained in the box along the ${\bf j}$ direction (which could be 0) and $L$ the lamellar period; this gives $L({\bf n}) = 1 \Big/ \sqrt{\sum_{\bf j} \left(n_j/L_j\right)^2}$.\cite{SymHH} For a fixed-length box as commonly used in canonical-ensemble simulations, both the lamellar orientation and its period are therefore discretized (i.e., $L$ can hardly be $L_0$); this problem is the most severe for cubic boxes.\cite{SymHH} To eliminate it, we change box lengths at a fixed $V$ (i.e., using rectangular boxes) with segmental positions re-scaled according to the Metropolis acceptance criterion. Compared to simulations in an isothermal-isobaric ensemble, this has the advantage that $\bar{\mathcal{N}}$ is constant during each simulation.

\begin{figure}[t]
\begin{center}
\includegraphics[height=5.5cm]{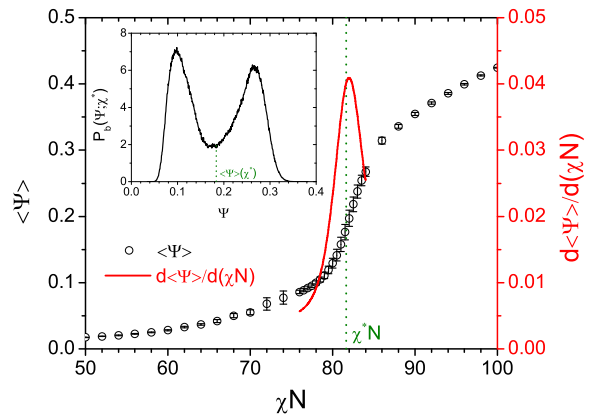}
\end{center}
\caption{\label{Psi}(Color online) Ensemble average of the order parameter, $\left<\Psi\right>$, and ${\rm d}\left<\Psi\right> / {\rm d} (\chi N)$. The error bar of $\left<\Psi\right>$ is taken as three times the standard deviation with sample correlations taken into account. The vertical line marks the ODT $\chi^* N$ determined using the equal-weight criterion\cite{Equal}, which is very close to the location of the maximum of ${\rm d}\left<\Psi\right> / {\rm d} (\chi N)$ (i.e., another way to determine ODT\cite{TransP}). The inset shows the reweighted histogram of $\Psi$ at $\chi^*$, $P_b(\Psi;\chi^*)$, where the vertical line marks $\left<\Psi\right>(\chi^*)$. $N=10$, $\sigma/a=0.3$, $N/\kappa=50$, and $\bar{\mathcal{N}}=10,000$.}
\end{figure}

We use a new order parameter for ODT, which characterizes the degree of positional order in lamellae. For a given direction ${\bf n}$, we calculate the volume fraction of A segments $\phi_{\rm A}(t) \equiv \hat{\rho}_{\rm A}(t) \big/ [\hat{\rho}_{\rm A}(t) + \hat{\rho}_{\rm B}(t)]$ as a function of position $t$ along ${\bf n}$ in a collected configuration (averaged over directions perpendicular to ${\bf n}$), and $\psi({\bf n}) \equiv \left|\int{\rm d}t \exp[4 \pi i t / L({\bf n})] f(t)\right| \big/ \int{\rm d}t f(t)$ with $f(t) \equiv 1 - |2\phi_{\rm A}(t) - 1|$ and $i \equiv \sqrt{-1}$. The order parameter $\Psi \in [0,1]$ for a collected configuration is defined as the largest $\psi$ over all possible lamellar orientations in the box. Fig.~\ref{Psi} shows the ensemble average of $\Psi$, $\left<\Psi\right>$, and ${\rm d}\left<\Psi\right> / {\rm d} (\chi N)$ vs. $\chi N$. We further use multiple histogram reweighting\cite{HR} to accurately locate ODT according to the equal-weight criterion\cite{Equal} described below.

The inset of Fig.~\ref{Psi} shows the histogram of $\Psi$ at a given $\chi$, $P_b(\Psi;\chi)$, with $\int_0^1 {\rm d}\Psi P_b(\Psi;\chi) = 1$. The double peak near ODT is a signature of the first-order phase transition. The ODT is determined as $\chi^*$ at which $\int_0^{\left<\Psi\right>} {\rm d}\Psi P_b(\Psi;\chi^*) = \int_{\left<\Psi\right>}^1 {\rm d}\Psi P_b(\Psi;\chi^*)$. To estimate the statistical error of $\chi^*$, we calculate ODT using the first- and second-half of our samples collected after equilibration, respectively, and take three times their largest deviation from $\chi^*$ determined using all the samples as the error bar. Note that, even with the same order parameter, there are other ways to determine the transition point\cite{TransP}, which give slightly different ODT for finite systems as shown in Fig.~\ref{Psi}. Systematic study of the finite-size effects requires much more extensive simulations and is out of the scope here.

\begin{table}[t]
\caption{\label{chiMF} Mean-field ODT $\chi^*_{\rm MF} N$.}
\begin{ruledtabular}
\begin{tabular}{cccc}
& $\sigma/a=0.3$ & $\sigma/a = 0.1 \sqrt{19}$ & $\sigma/a = 2/\sqrt{3}$ \\
\hline
$N=10$ & 10.047 & & 11.427 \\
$N=20$ & 10.405 & 10.462 & 11.102 \\
\end{tabular}
\end{ruledtabular}
\end{table}
With the same Hamiltonian as used in our FOMC simulations, we determine the mean-field ODT $\chi^*_{\rm MF}$ using the random-phase approximation\cite{FOMC}. Table~\ref{chiMF} lists $\chi^*_{\rm MF} N$ for various $N$ and $\sigma/a$. We note that the mean-field results are independent of $N/\kappa$ and $\bar{\mathcal{N}}$, and can be understood based on our previous work\cite{FRI,FOMC}.

\begin{figure}[t]
\begin{center}
\includegraphics[height=5.5cm]{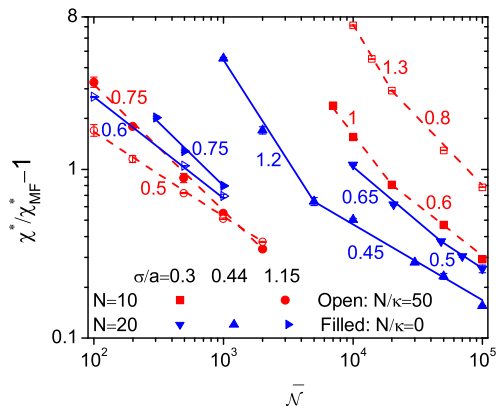}
\end{center}
\caption{\label{ODT}(Color online) Our results on the ODT shift. The negative slope $k$ is given next to each line. See text for details.}
\end{figure}

Collecting all the data, Fig.~\ref{ODT} shows our results of ODT shift vs. $\bar{\mathcal{N}}$ at various $N$, $\sigma/a$ and $N/\kappa$.\cite{Data} We find $\chi^* / \chi^*_{\rm MF} - 1 \propto \bar{\mathcal{N}}^{-k}$ in all the cases. While this is consistent with the functional form of FH prediction, our ODT shift is larger than their prediction for all the cases we have studied. We also find a decrease in the negative exponent $k$ around $\bar{\mathcal{N}}_{\rm cp} = 2 (N-1)^3 \big/ N^2 (\sigma/a)^6$ in the cases of $\sigma/a=0.3$ and $0.1\sqrt{19} \approx 0.44$, which corresponds to the FCC close packing of polymer segments as hard spheres. This is therefore due to the local packing of segments (i.e., the short-range correlations). Denoting the $k$-value for $\bar{\mathcal{N}} < \bar{\mathcal{N}}_{\rm cp}$ by $k_1$ and that for $\bar{\mathcal{N}} > \bar{\mathcal{N}}_{\rm cp}$ by $k_2$, we find that $k_1 - k_2$ decreases with increasing $N$ (i.e., in the case of $\sigma/a=0.3$), which leads to decreasing $\sigma$ (thus weaker correlations) at constant $R_{e,0}$. One may therefore expect a single power-law decay of $\chi^* / \chi^*_{\rm MF} - 1$ with increasing $\bar{\mathcal{N}}$ in the limit of $N \to \infty$ (i.e., $\sigma \to 0$), which has no such correlation and is the case studied by Fredrickson and Helfand\cite{FlucFH}. On the other hand, increasing $\sigma/a$ at constant $N$ leads to stronger correlations and thus increasing $k_1 - k_2$ (i.e., in the case of $N=20$). We also note that varying $N$ at constant $\sigma/a$ exhibits both the correlation and chain discretization effects. To examine the latter alone, one may compare the case of $N=10$ and $\sigma/a=0.3$ with that of $N=20$ and $\sigma/a=0.1\sqrt{19}$ (both of which have $\sigma = 0.1 R_{e,0}$). We expect that the chain discretization effects diminish quickly with increasing $N$, as supported by Fig.~6(a) of Ref.~\cite{FOMC} showing how $\chi^*_{\rm MF} N$ varies with $N$.

In the case of $\sigma/a = 2/\sqrt{3} \approx 1.15$, $\bar{\mathcal{N}}_{\rm cp} < 15$ for both $N$-values, and we did not perform simulations at $\bar{\mathcal{N}} < 100$. This large $\sigma/a$-value also makes simulations at $\bar{\mathcal{N}} > 2000$ expensive to do. Nevertheless, we see from Fig.~\ref{ODT} that increasing $N$ increases the ODT shift in this case, in contrast to that of $\sigma/a = 0.3$. On the other hand, as we increase $N/\kappa$ from 0 to 50, $k_2$ decreases here and the two lines with the same $N$ in Fig.~\ref{ODT} cross at $\bar{\mathcal{N}}_{\rm cr} \approx 1300$ (for $N=10$) or 2500 (for $N=20$). Therefore, $\chi^* N$ decreases for $\bar{\mathcal{N}} < \bar{\mathcal{N}}_{\rm cr}$ and increases for $\bar{\mathcal{N}} > \bar{\mathcal{N}}_{\rm cr}$ with increasing $N/\kappa$. While the latter is consistent with the case of $\sigma/a=0.3$, we find that both $k_1$ and $k_2$ increase with increasing $N/\kappa$ there.

To summarize, we have performed extensive FOMC simulations with experimentally accessible fluctuations to systematically and  unambiguously quantify the ODT shift of symmetric DBC from the mean-field prediction. Exactly the same model system (Hamiltonian) is used in both our simulations and mean-field theory; the ODT shift is therefore due to the fluctuations/correlations neglected by the latter. We hope this work will stimulate the development of advanced theories better describing the fluctuation/correlation effects identified here.

We thank Prof.~Marcus M\"uller, Prof.~Mark Stoykovich, and Dr.~Kostas Daoulas for helpful discussions. This work was supported by the U.S. Department of Energy, Office of Basic Energy Sciences, Division of Materials Sciences and Engineering under Award DE-FG02-07ER46448.

\bibliographystyle{unsrt}

\end{document}